\documentclass{article}
\usepackage{amssymb,amsmath,amsthm}
\theoremstyle{plain}

\theoremstyle{definition}

\makeindex

\begin{document}

 \begin{center}
  Generalized Shock Model Based On The Frequency of Shocks: A Simple Approach \\
  Viswanathan Arunachalam \footnote{Corresponding author's email:  varunachalam@unal.edu.co}  \\
  Department of Statistics \\    Universidad Nacional de Colombia \\ Bogot\'a, Colombia.
\end{center}
\begin{abstract}
In a $\delta-$shock model, a system subject to randomly occurring shocks, the system fails when the time between two successive shocks lies below a threshold $\delta$. In this note, we  study the generalization of this model where such  $\delta-$shocks are accumulated and the system fails on the occurrence of $k^{th}$ such a $\delta-$shock. The probability distribution of the system failure time and the statistical characteristics are explicitly obtained. Normal approximation  to the failure time distribution is proposed.
\end{abstract}

  \begin{center}
stochastic model;   $\delta-$shock model ; frequency dependent;  Reliability modeling
\end{center}


\section{Introduction}

   An  illuminative way of modeling deteriorating systems is through the use of shock models. Shocks are random events which  cause  certain damage to the system leading to its deterioration and are assumed to be additive. The system fails when the accumulated damage crosses a threshold. However Lam \cite{Lam} and Rangan and Tansu\cite{RT} have considered $\delta-$ shock models which  concentrate on the frequency of shock occurrences, as contrasted to  the accumulated damage of the earlier models. In these class of models, system fails when two successive shocks are not separated by a sufficiently long interval $\delta$ (which could be random). Thus any shock is considered to be a lethal shock leading to system  failure if the time between this shock and the previous shock
 is less than $\delta$.  The purpose of using $\delta$ as the threshold to failure is to model the recovery time of the system from shocks.  It is eminently possible for systems to successfully withstand a few of these lethal shocks before failure. For instance Eryilmaz\cite{Ery} recently proposed   run-related generalization of $\delta-$ shock  model such that the system fails when k consecutive inter arrival times are less than a threshold $\delta$, where $\delta $ is constant. The purpose of this paper is to generalize the $\delta-$ shock models to allow the system to accumulate $(k-1)$ such shocks and derive the failure time distribution and its statistical characteristics.  The $\delta-$ shock models have  many applications in various fields from system reliability  to  neuronal firing models(\cite{Abd}, \cite{Aru}, \cite{Lam2}, \cite{Lam3}, \cite{RG}).
\section{The Model}

 A new system which is put on operation at $t=0$ is subject to randomly occurring  shocks. The interval between shocks are assumed to be independently and identically distributed random variables with distribution function $F(\cdot)$. A shock is classified as potentially lethal shock if the time elapsed from the previous shock  to this shock is less than a certain threshold  $\delta$. The threshold $\delta$ is a random variable with distribution function $G(\cdot)$. The shock arrival times and threshold times are assumed to be independent of each other. The system can survive $(k-1)$ such potentially lethal shocks and system failure occurs at the instant of $k^{th}$ such  shock, where $k$ can be any positive integer greater than 1. Our interest is in computing the probability distribution of $W$, the random variable representing time to failure of the system and its statistical characteristics.

 We note that during $W$ a random number  of $N$ of shocks occur of which exactly $(N-k)$ of them are not potentially lethal shocks and     $k$  are potentially lethal shocks, the $k^{th}$  shock is to occur leads to system failure. Thus $W$ comprises of the sum of a random number of $N$ intervals of which  $(N-k)$ of them are greater than $\delta$ and $k$   are less than $\delta$. We define a sequence of independently and identically random variables $X_i'$s  which are distributed as $Z$ but conditional on $Z>\delta$. We also define a  sequence of independently and identically distributed random variables $Y_i'$s  which are distributed as Z but conditional on $Z\leq \delta$.  With above definitions, $W$ can be represented as
 \begin{equation} \label{eq1}
 W =  \sum_{i=1} ^{N-k} X_i + \sum_{i=1} ^{k}Y_i \ .
 \end{equation}
The total number of terms $N$ in the summation from the assumptions of the model, follows a negative binomial distribution given by

 \begin{equation} \label{eq2}
 P(N=n) = {n-1\choose k-1} p^k q ^{n-k} ~, ~~~~n=k,k+1,k+2,\ldots
 \end{equation}
where $p = P(Z \leq  \delta)$ and $p+q = 1$.

We define the conditional distributions of $X_i$ and $Y_N$ as

 \begin{equation} \label{eq3}
\alpha (t) = P\left( t < Z < t+dt  \mid Z  >  \delta \right) =
\frac{f (t) G  (t)}{P(Z > \delta)}
 \end{equation}

 and
  \begin{equation} \label{eq4}
\beta (t) =   P\left( t < Z < t+dt  \mid Z \leq \delta \right) =
\frac{f (t) \overline{G(t) }}{P(Z \leq \delta)} \ .
 \end{equation}
Now  $h(t)$ the probability distribution  $W$ is obtained as

\begin{eqnarray} \label{eq5}
h (t) & = &     P\left( t < W < t+dt \right)  \nonumber \\ & = &
\sum_{n=1} ^{\infty} P\left( t < W < t+dt \mid N = n \right)
P(N=n) \nonumber
\\
 & = & \sum_{n=1}
^{\infty}{n-1 \choose k-1}  \left(\alpha^{(n-k)}  \ast \beta^{(k)} (t) \right) \left[P(Z \leq \delta)\right]^k \left[P(Z>
\delta)\right]^{n-k}
 \end{eqnarray}

 where $\alpha^{(n-k)}  \ast \beta^{(k)} $ is the convolution of
$k$- fold convolution of $\alpha (t)$ with $(n-k)$ fold convolution of $\beta (t)$.  Taking the Laplace transform on both sides of (\ref{eq5}) and using (\ref{eq3}) and (\ref{eq4}) we obtain

 \begin{equation} \label{eq6}
L_h (s) = \left( \frac{L_{f\overline{G}}(s)}{1- L_{fG} (s)} \right) ^ k \ .
 \end{equation}
where $L_{f\overline{G}}(s)$ and $L_{fG} (s)$ are the Laplace
transforms of the functions  $f (t) \overline{G(t)}$ and
$f (t)  G (t)$, respectively. Given the specifications of
the distributions $F$ and $G$, one might be able to invert
(\ref{eq6}) to obtain the probability density function $h(t)$.  In
cases where a closed form inversion of $L_h (s)$ is not possible,
one can use the algorithms proposed by Abate and Whitt\cite{Aba}
  for numerically inverting Laplace transforms
which are designed especially for probability density functions.

The moments of $W$ for any shock arrival distribution $f(t)$ and threshold distribution $g(t)$ are obtained by
  differentiating $L_h (s)$ with respect
to $s$ and setting $s = 0$. It can be easily shown after some algebra that

 \begin{equation} \label{eq7}
E(W) =  k \frac{E(Z)}{P(Z \leq \delta)}  = k \mu \ ,
 \end{equation}
  and

\begin{equation} \label{eq7a}
 Var(T)=  k\left( \frac{E(Z^2)}{P(Z \leq \delta)}+\frac{2E(Z)E(Z\mid Z
>\delta)P(Z > \delta)-E^2 (Z)}{P(Z \leq
\delta)^2} \right) = k\sigma ^2  \ .
 \end{equation}

At this juncture we wish to observe that the results of Lam \cite{Lam} and Rangan and Tansu\cite{RT} are reduced  by setting $k=1$ in our model in accordance with their model assumptions.

We now present an example by considering the lifetime $\delta$ to be a
constant to illustrate our model. Let us first assume that the
potentially lethal shock arrive according to exponential density $f(t) = \lambda
e^{-\lambda t}$  and the threshold distribution
\[
 G_{\delta} (t) = \left\{ \begin{array}{ll}
                     0, & 0 \leq  t < \tau  \\
                     1, &  t \geq \tau
                      \end{array}
                      \right. \ .
 \]

 Equation (\ref{eq6}) in this case reduces to
 \begin{eqnarray*} \label{eq6a}
L_h (s) & = & \left(\frac{\lambda }{s+\lambda }\right)^k \frac{[1-e^{-(s+\lambda)\tau}]^k}{[1-\frac{\lambda }{s+\lambda} e^{-(s+\lambda)\tau} ]^k} \\
 & = & \sum _{j=1} ^{\infty} \sum _{i=1} ^k  (-1)^i {k \choose i}{j+k-1 \choose j} \left(\frac{\lambda }{s+\lambda }\right) ^{j+k}   e^{-(s+\lambda)(j+i)\tau} \ .
 \end{eqnarray*}
 Inverting  the  above Laplace transform, we get density function of  $W$, as
  \begin{eqnarray} \label{pdf}
h(t) &=&   \lambda \frac{ \lambda ^{k-1} e^{-\lambda t}}{(k-1)!} \sum _{j=0}^{\infty} \sum _{i=0}^{k}   (-1)^i {k \choose i} \frac{\lambda ^{j}}{j!}\left[(t-(j+i)\tau) U(t-(j+i)\tau) \right] ^{(j+k-1)}  \ .
 \end{eqnarray}
 where $U(t-c)$ is the Heaviside unit step function
 \[
 U (t-c) = \left\{ \begin{array}{ll}
                     0, & 0 \leq  t < \tau  \\
                     1, &  t \geq \tau
                      \end{array}
                      \right. \ .
 \]

From  Equations (\ref{eq7}) and (\ref{eq7a}), $E(W)$  and $Var(W)$ respectively are
 \begin{equation} \label{eq9}
E(W) = k \frac{1}{\lambda (1- e^{-\lambda \tau})}.
 \end{equation}
 \begin{equation} \label{VW}
Var(W) =  k \frac{1+2\lambda \tau e^{-\lambda \tau}}{\lambda^2 (1- e^{-\lambda\tau})^2} \ .
 \end{equation}


 As a second example, if the stimuli arrival
 distribution is uniform so that
\[
f(t) = \frac{1}{b-a}~~~~~~~~~ a< t <b,
\]  and constant lifetime  $\tau$ then we can derive

\begin{equation} \label{eq9}
L_h (s) =\left( \frac{e^{-sa} - e^{-s\tau}}{s(b-a)-e^{-s\tau} +e^{-sb}} \right)^k
 \end{equation}
 and
 \begin{equation} \label{eq10}
E(W) = k\frac{b^2-a^2}{2(\tau -a)} \ .
 \end{equation}
The variance of $W$ is given by
 \begin{equation} \label{cvTU}
Var(W) = k \frac{2\mu _2  (\tau -a)+\mu _1 (b^2 -2 \tau ^2 + a^2)}{2\mu_1 (\tau-a)}
 \end{equation}
where $\mu _1$ and $\mu _2$ are first and second raw moments of  $f(t)$.

 \section{ Normal Approximation}
 A closer look at Equation (\ref{eq6}) reveals that  the time for system failure $W$, is sum of the $k$ independently and identically distributed random variables $S_1, S_2,\ldots,S_k$, where each $S_i$ is the time  between two successive potentially lethal shocks.  Also the Laplace transform of the probability distribution of each $S_i$ is given by
 $$ \frac{L_{f\overline{G}}(s)}{1- L_{fG} (s)} \ .
 $$
  The mean $\mu$ and variance $\sigma ^2$ are  then specified by the Equations (\ref{eq7}) and (\ref{eq7a}). Now for   large, we can  invoke central limit theorem, so that  $h(t)$ can be approximated by the normal distribution
   \begin{equation} \label{N2}
h(t)=\frac{1}{ \sigma \sqrt{2k \pi}}exp^{-\frac{1}{2k\sigma ^2} (t-k\mu)^2}  \ .
 \end{equation}

 Since  the quantity of interest in practical applications is the time for the  first crossing of the $k^{th}$ potentially lethal  shock, the equation(\ref{N2}) will be very useful in applications.
 \section{Conclusion}

 Shock models are versatile in terms of  applications to diverse areas from fatigue failure of materials to neuron firing in neurophysiology. Thus any useful contribution in such models will helpful its  understanding of  system failure.  As the model assumes that any potentially lethal shock is stored in the system and the system fails when the number of stored  potentially lethal shocks reaches k.  A more interesting problem arises if it is assumed that each shock has a random lifetime $\delta$ and can not be stored for more than $\delta$ units of time and $k$ such shocks are needed for system failure.



\end{document}